\documentstyle[preprint,tighten,eqsecnum,epsfig,aps]{revtex}
\input{psfig}

\def\beq{\begin{equation}}   \def\eeq{
\end{equation}}

\begin{document}
\title{ Coulomb scattering of quantum dipoles in QED.
  } \author{ B. Blok\thanks{E-mail: blok@physics.technion.ac.il} }
\address{Department of Physics, Technion -- Israel Institute of
Technology, Haifa 32000, Israel}
\maketitle

\thispagestyle{empty}

\begin{abstract}
We calculate the total scattering cross-section  of a dynamical
quantum electrically neutral dipole in QED of the infinitely heavy
charge and of the infinitely heavy  dipole in the leading order in
$\alpha_{em}$.
 \end{abstract}

\pacs{} \setcounter{page}{1} \section{Introduction}
\par The study of the space-time evolution of quantum dipoles in QED and
QCD attracted a lot of attention recently
\cite{Farrar,FS13,FS,FS1,M,Gribov,Blok,Blok1}.
\par In particular in a recent paper \cite{Blok1} we have showed
that the evolution of the approximately symmetric quantum dipole
can be described using the renormalized wave function. The purpose
of the present paper is to use the formalism developed in ref.
\cite{Blok1} to calculate the cross-section of a quantum dipole
off the Coulomb center and off the other dipole.
\par Our main result is the use of the regularized dipole wave
function of ref. \cite{Blok1} for the precise
definition of the dynamical dipole time-dependent cross-section and it's
calculation. We obtain
\beq
\sigma(T) =g^2(S(T)/2)\log(E/\delta)
\label{1a}
\eeq
for the scattering of the approximately symmetric dynamical dipole with  the
center of mass energy E of the infinitely heavy Coulomb center, and
\beq
d\sigma(T)/dt=g^2S(T)b^2/2
\label{1b}
\eeq
for the scattering of the dynamical dipole of the infinitely heavy dipole.
Here $S(T)$ is the transverse area of the dynamical dipole at time T
after its creation, and b is the transverse radius of the target dipole.
\par Our results are in agreement with the ones by
D.Soper and J.Gunion
\cite{GS,Farrar} that
the scattering cross-section off the dipole is proportional to $b^2$,
and the well known statement that the dipole cross-section of the
charged Coulomb center increases logarithmically with energy
\cite{D,LL1,AB}. In the leading order our cross-section coincides with
the one obtained in ref. \cite{Farrar} in a different way.
\par The paper is organized in the following way. In the second chapter
we discuss the general formalism of calculations. In the third
chapter we carry the calculations and our results are summarized
in conclusion.

\section{General Formalism}

\subsection{Dipole way function.}

\par It was explained in ref. \cite{Blok1} that one can describe the
dipole evolution by means of the $a,b\rightarrow 0$ limit of the dipole
wave function
\begin{eqnarray}
\Phi(\vec r,\vec R,t)&=&(1/\sqrt{2E})\sqrt{2a/\pi}\exp(i\vec Q\vec
R)\exp(iEt)\int
d^2q_t\exp(-(a+ib)(\vec q_t-\vec
k_t)^2\nonumber\\[10pt]
&+&2(it/(E))q^2_t)\nonumber\\[10pt]
&\times&\exp(i\vec q_t\vec \rho ),\nonumber\\[10pt]
\label{kj9}
\end{eqnarray}
(where we neglect the frozen evolution in z direction). In
coordinate space one has
\begin{eqnarray}
\Phi(r,R,t)&=&(1/\sqrt{2E})\exp(i\vec Q\vec R)\exp(iEt)
\frac{\sqrt{2a\pi}}{a+i(b+4t/E)}\nonumber\\[10pt]
&\times&\exp{(-(\vec \rho -4\vec k_tt/E)^2/((a+i(b+4t/E))+i\vec
k_t(\vec \rho-4\vec k_tt/E)}). \nonumber\\[10pt]
 \label{34} \end{eqnarray}
 For this dipole we
have time dependent radius: \beq <\rho^2(t)>=a/2+b^2/2a+8(b/a)\hbar
t/E+8t^2/(E^2a) \label{j34} \eeq \beq \rho^i(t)=4k^it/E.
\label{j35} \eeq
\par Our strategy will be to use these wave functions in order to calculate
scattering amplitudes. We shall limit ourselves here with the most
interesting case of quantum dipole ($\vec p_t=0$).

 \subsection{Scattering amplitude calculation.}

\par It was explained in ref. \cite{Gribov1} ( see also the discussion
relevant to the given case in ref. \cite{Blok1}) that in order to calculate
the amplitude, one can use casual ordering of the old perturbation theory
\cite{Grandy}.
All other contributions are suppressed by the powers of
energy.
\subsection{Time dependent cross-section.}
\par We are interested in the full cross-section that is proportional
to
\beq
\partial P(T)/\partial T
\label{3} \eeq Here P(T) is the probability of all possible
transitions in time T \beq P(T)=\sum_f M_{if}M_{if}^* \label{4}
\eeq where i is our initial dipole configuration, and $f$ is a
full system of intermediate states. Since we carry our
calculations using free wave packets we can take as a full system
the free particle wave functions: \beq \psi(\vec r_1,\vec
r_2)=\frac{1}{\sqrt{\epsilon_{p_1}\epsilon_{p_2}}}\exp(i\vec
k_1\vec r_1)+i\vec p_2\vec r_2)+i\epsilon_{p_1}t-i\epsilon_{p_2}t
\label{3a} \eeq where $\epsilon_p=\sqrt{p^2+M^2}$. Then the
summation over the intermediate states reduces to integration over
$d^3p$. The time-dependent cross-section is defined as \beq \sigma
(T)=\partial P(T)/\partial T \label{5} \eeq Here the wave
functions of the dipole constituents must be normalized like
$1/\sqrt{V}$, corresponding to one particle in a unit volume.

\section{Calculation of the scattering cross-section}

\subsection{Scattering of the infinitely heavy Coulomb center.}

\par In order to calculate the scattering cross-section we need to calculate
the difference of transition amplitudes from the initial dipole to a
$\vec p_1\vec p_2$ intermediate state, take a square and then integrate over
$d^2p_1d^3p_2$.
It is easy to see that the corresponding amplitude is just the difference
\begin{eqnarray}
M&=&\sqrt{2a/\pi}\int
ds^T_0(\exp((a(p-k)^2+2i(2k_tp_t+k^2_t)t/E\nonumber\\[10pt]
&-&
\exp(-(a(p+k)^2+2i(-2k_tp_t+k^2_t)t/E)\nonumber\\[10pt]
\label{5a}
\end{eqnarray}
Here $\vec p=\vec p_1-\vec p_2$ and $\vec k$ is a transferred
momenta during the scattering. The first term in the latter
expression corresponds to the
Coulomb
photon emission from a
positive charged component of the dipole (a particle) while the
second-from the negative charge one. The square of this expression
is given by the difference $M_1-M_2$, corresponding to direct and
cross product, i.e. in the first term we consider the product of
amplitudes corresponding both to particle, or both to
antiparticle, and the second term corresponds to the product when
one of the amplitudes corresponds to particle and the other to
antiparticle. It is easy to obtain
 \begin{eqnarray}
M_1&=&\int^dt\int dt'
\int d^2p_t\exp(2i(p_t+k)^2(t-t')-2ip_t^2(t-t'))/E)+ik_3(t-t'))\nonumber\\
[10pt]
\\&\times&\exp(-(a+ib)(p_t-k)^2+(p_t+k)^2)\sqrt{2a/\pi}\nonumber\\[10pt]
\label{6}
\end{eqnarray}
The second diagram
is given by
 \begin{eqnarray}
M_2&=&\int^T_0dt\int dt'
\int
d^2p_t\exp(i(p_t+k_t)^2-p^2_t)t/E-i((p_t-k_t)^2-p^2_t)t'/E)\nonumber\\[10pt]
&+&ik_3(t-t'))
\exp(-(a+ib)(p_t+k_t)^2-(a-ib)
(p_t-k_t)^2)(2a/\pi)/(k^2+\delta^2)^2d3k\nonumber\\[10pt]
\label{7}
\end{eqnarray}
\par We can carry the integration over the transverse momentum $p_t$.
We obtain: \beq M_1=\int
dtdt'\exp(2a-2k^2_ts^2/(E^2a)+ik^2_ts/E+ik_3s/(k^2+\delta^2)^2+h.c.
\label{8} \eeq \beq M_2=\int dt\int dt'
\exp(-2(4(t+t')/E+b))^2k^2_t/(a)+ik^2_t/E+ik_3s)+h.c. \label{9}
\eeq We now differentiate $M_1$ and note that if $T\ll
\sqrt{E^2a}/k_t$, we can neglect the real term in the argument of
the exponent in eq. \ref{8} . The integration over s then gives
the result that is peaked near $0=k_3+k^2_t/E$. We can
approximately substitute the result of this integration by a
$\pi\delta (k_3+k^2_t/E$, and then carry the integration over
$k_3$. The net result since $k_t\ll E$, is that we can put
$k_3=0$. Consider now the eq. \ref{9}. There we can put after
differentiation approximately $u=T+t'\sim 2T$, while the
integration over s and $k_3$ as before gives us $k_3=0$. Thus we
obtain \beq \sigma(T)\equiv \partial (M_1-M_2)\partial T=2\pi\int
d^2k_t(2a-\exp(2(4iT/E+ib)^2k^2_t)/a)/(k^2_t+\delta^2)^2
\label{10} \eeq for times much smaller than the coherence length
we can expand the exponents in eqs. (\ref{9}-{10}). and obtain
\beq \sigma(T)=4\pi\int d^2k_t
(a^2+((4T/E+b))^2k^2_t/a)/(k^2_t+\delta^2)^2 \label{11} \eeq

\subsection{Renormalization.}

\par The latter calculation is however not an end of the story. It was
explained in ref. \cite{Blok1} that we must now go to the limit
$a,b\rightarrow 0, 8a/b\rightarrow D, E^2a\rightarrow v^2_t$
to keep gauge invariance of an amplitude.
Then one immediately obtains \beq \sigma(T)=\pi\int d^2k_t/k^2_t\times
(DT/E+v^2_tT^2) \label{13} \eeq Here $D$ is the diffusion
coefficient ($D=2$ for the longitudinal vector current
\cite{Farrar,Doc}). In other words the total cross-section is just
\beq \sigma(T)=S(T)\int dk_t/k_t \label{14} \eeq integrating the
latter equation with logarithmic accuracy we obtain using
$k_t=2E\sin (\theta/2)$ \beq P(T)=S(T)\log(E/\delta) \label{15}
\eeq Here $S(T)$ is the dipole transverse area at time T,
calculated in ref. \cite{Farrar,Blok1}: \beq S(T)=DT/E+v^2_t
\label{16} \eeq It is now clear what is the relevant time scale:
\beq T\ll 1/(v_tk_t)=E/(p_tk_t) \label{17} \eeq Here $v_t=p_t/E$
is a renormalized characteristic transverse momentum of the
dipole.

\subsection{Dipole-dipole scattering.}

\par In the previous section we carried the calculation
for the scattering on the infinitely heavy
Coulomb
center. Exactly the same calculation can be made for the dipole-dipole
scattering on the infinitely heavy dipole with the transverse scale b. It
is straightforward to see that the dipole contributes the multiplier
\beq \vert 1-\exp(i\vec k_t\vec b)\vert^2=4\sin(\vec k_t\vec
b/2)^2 \label{17a}\eeq For small $b$ we can expand the latter
sinus in Taylor series and obtain \beq \sigma(T)=\sigma (T)b^2\int
dk_t k_t \label{18} \eeq We can move from integration over
$k_t=2E\sin(\theta/2)$ to integration over the invariant kinematic
variable $t=k^2_t$. Then \beq \frac{dP(T)}{dt}=\sigma(T)b^2/2
\label{19} \eeq
 We see that the differential cross-section of dipole-dipole scattering
does not increase with energy, while the full cross-section must
be determined through the integration over the allowed kinematic range of
$t$. This is in contrast to the scattering on the infinitely
heavy coulomb center,
where the full cross-section increased logarithmically with energy.

\par Our result \ref{19} in the leading order coincides with the one by
Farrar et al \cite{Farrar} (see also ref. \cite{Doc}) obtained in a
different way.

\section{Conclusion.}

\par In this paper we have shown that using the dipole wave functions
defined
in ref. \cite{Blok1} and old perturbation theory \cite{Grandy}, it is
possible to calculate the scattering
cross-section of the dipole of the infinitely heavy dipole and of
the coulomb center.
\par Our main result is the precise definition of the dynamical dipole
time-dependent cross-section as a time-dependent matrix element between
the wave functions of ref. \cite{Blok1} and it's calculation: \beq
\sigma(T) =g^2S(T)\log(E/\delta)
\label{1aa}
\eeq
for the scattering of the dynamical dipole of the infinitely heavy
Coulomb center, and
\beq
d\sigma(T)/dt=g^2S(T)b^2/2
\label{1ba}
\eeq
The result coincides in the leading order with the one due to ref.
\cite{Farrar} that was obtained in a different way. The dependence on
the target dipole size for the scattering of the heavy dipole and
on the energy coincide with the known general results due to
\cite{GS,D,LL1,AB}.
\par Our results can be easily extended to the scattering of the dynamical
dipole in the strong field. The only difference is that instead of
a full system of the plain waves used here one must use the the
full system of the wave functions in this external field.
\acknowledgements{The author thanks prof. L. Frankfurt for reading
a manuscript and numerous useful discussions.}

\end{document}